\documentclass{cargese}\usepackage{graphicx}

\let\footnote\savefootnote
\let\footnotetext\savefootnotetext 
\let\footnoterule\savefootnoterule 
\normallatexbib

\def\np#1#2#3{Nucl. Phys. {\bf{B#1}} (#2) #3}

\begin{document}

\articletitle[The NS five-brane partition sum]
{The NS five-brane partition sum}

\author{Marcel Vonk}

\affil{Institute for Theoretical Physics, University of Amsterdam \\ and \\
Spinoza Institute, PO Box 80195, 3508 TD Utrecht, The Netherlands}    

\email{m.l.vonk@phys.uu.nl}

\begin{abstract}
We calculate the euclidean partition function of the type IIA NS five-brane
wrapped on an arbitrary Calabi-Yau space in a double-scaling decoupling limit
and in the presence of a flat RR 3-form background field. The result is the 
product of a theta function, coming from the classical fluxes of the self-dual 
tensor field, and a factor representing the quantum contributions. The 
quantum factor turns out to be related to topological B-model string amplitudes, 
and both factors satisfy a holomorphic anomaly equation. The result can
teach us more about little string theories and about instanton corrections to
four-dimensional effective quantities.
\end{abstract}

The work reported here was done in collaboration with R. Dijkgraaf and E.
Verlinde \cite{dvv02}. This text is intended to give a brief overview of this
work. Only the main results are reported here; the calculations and many other
details can be found in the original paper.

\section{Description of the system}
We consider euclidean type IIA string theory, compactified on an arbitrary
Calabi-Yau manifold $X$. Around $X$, we wrap one or more euclidean NS 
five-branes, so that these objects are pointlike in the remaining four 
directions. We turn on a background RR-field $C$ which satisfies $dC=0$ along
the directions of $X$. Finally, we take a double scaling limit where we send
both the distance between the five-branes and the string coupling constant to 
zero in such a way that their ratio becomes infinite.

The reason to study this system is twofold. First of all, the world-volume
theory on an NS-five-brane is the so-called little string theory, which is still
not completely understood. As shown by Giveon and Kutasov \cite{gk99}, in the
above double scaling limit this theory turns into a weakly coupled field theory.
Therefore, the calculation of the partition function of this system could teach 
us something about little string theory. Secondly, in Calabi-Yau
compactifications there are instanton corrections to several four-dimensional
quantities, coming from wrapped NS five-branes. Our computation could be helpful
in determining these instanton corrections.

\section{The classical factor}
The bosonic part of the low-energy field theory on the NS five-brane contains a 
self-dual two-form field and five scalar fields. For clarity, we will not take
the fermions -- which are completely determined by supersymmetry -- into
consideration here. Also, the scalar fields do not contribute to the classical
partition function. Four of them, corresponding to the position of the NS
five-brane, are fixed while the fifth one cannot have any winding numbers since
the Calabi-Yau does not have any closed one-cycles. Therefore, we want to study
the classical partition sum of a self-dual two-form $B$ coupled to a flat 
background three-form field $C$. Ignoring the subtleties of the self-duality, 
this system can be described by the action
\begin{equation}
 S = \frac{1}{4 \pi} \int_X \frac{1}{2} (H - C) \wedge * (H - C) - i H \wedge C,
\end{equation}
where $H=dB$ is the field strength of the two-form field.

The field $H$ has quantized fluxes around the conjugate three-cycles $A^I$,
$B_I$ of the Calabi-Yau:
\begin{equation}
 \int_{A^I} H = 2 \pi n^I, \qquad \int_{B_I} H = 2 \pi m_I.
\end{equation}
Similarly, we denote the (fixed) periods of the $C$-field around these cycles 
by $x_A^I, x_{B,I}$, and the fluxes of the holomorphic $(3,0)$-form on $X$
around the $A$-cycles (defining its complex strcture) are denoted by $z^I$. 
Note that the variables $x$ and $z$ are not quantized.

The partition sum is now a sum over the fluxes $m,n$, which can be evaluated 
using a Poisson resummation. It is well-known how to incorporate self-duality 
in such a calculation \cite{witten96}: basically, one has to choose a spin 
structure $(\alpha^I, \beta_I)$ on the manifold and take a holomorphic root of 
the resulting expression. The result of this somewhat lengthy but 
straightforward calculation of the classical partition function is
\begin{equation}
 Z^{cl}_X = \overline{\Theta_{\alpha, \beta} (x^I;z^I)}
 \label{eq:classicalresult}
\end{equation}
i.\ e.\ a modified theta-function depending on the background field fluxes $x^I$
and on the periods $z^I$. The sum over the fluxes $m$ and $n$ (only one sum is
left because of the holomorphic root) has turned into the usual sum inside the 
theta function. For the full expression for $\Theta$, the reader is referred to
\cite{dvv02}.

\section{T-duality}
It is known that after a transversal T-duality, the NS five-branes (which are
four-dimensional instantons for the $B_{\mu \nu}$ field), turn into 
gravitational instantons. More specifically, it has been shown in \cite{ov96} 
that the T-dual system in four dimensions is a Taub-NUT space (without 
five-branes) which in the limit where we let the compactification circle grow 
to infinite size becomes an ALE-space of type $A_{k-1}$, where $k$ is the 
number of five-branes in the original picture.

Comparing the limits on both sides of the T-duality, we find that the double
scaling limit corresponds to a weak coupling limit with small Planck length, i.\
e.\ we can use an ${\cal N} = 2$ supergravity approximation on the type IIA side. The three-form
field on the Calabi-Yau obtains a leg in the four remaining directions after the
T-duality, so for every three-cycle $A^I$ on the Calabi-Yau, we obtain a vector
field with field strength $F^I$ in the supergravity limit. (The cycles $B_I$
correspond to the Hodge duals of these forms.)

\section{The quantum factor}
In the ${\cal N}=2$ supergravity theory, we can again calculate the classical
partition function of a single five-brane by summing over the fluxes of the
gauge fields $F^I$, and one finds exactly the result (\ref{eq:classicalresult}).
However, the interesting thing is that on this side one can also calculate the
quantum contributions to the partition function. In fact, it was shown in
\cite{agnt94} that the supergravity action obtains quantum corrections of the
form
\begin{equation}
 S^{qu} = \int \sum_{g=1}^{\infty} R_- \wedge R_- (g_s T_-)^{2g - 2} {\cal
 F}_g(z, \bar{z}),
\end{equation}
where $R_-$ is the anti-self-dual part of the curvature, $T_-$ is the
anti-self-dual part of the graviphoton field strength (which is one of
the gauge field strengths in our description above), and the ${\cal F}_g$ are
topological string amplitudes at genus $g$.

A straightforward calculation shows that the integral over the curvature gives a
factor of $k-1$ -- corresponding to the number of five-branes -- and $g_s$ and
$T_-$ in our scaling limit combine into an effective coupling constant
$\lambda$, so the final result for the partition function of a single five-brane
is
\begin{equation}
 Z_X = \overline{\Theta_{\alpha, \beta} (x^I;z^I)} \exp \left( 
 \sum_{g=1}^{\infty} \lambda^{2g - 2} {\cal
 F}_g(z, \bar{z})\right).
 \label{eq:fullresult}
\end{equation}

\section{Open problems}
Two important open questions arise from our calculations. First of all, since
$k$ five-branes correspond to an $A_{k-1}$-singularity, it is not clear what
happens in the case of a single five-brane. In fact, the result we get for $k$
five-branes in the Coulomb phase on the supergravity side seems to be the 
$(k-1)^{th}$ power of the result (\ref{eq:fullresult}). In our paper
\cite{dvv02}, we give some arguments for the fact that there is indeed a 
``missing five-brane'' at infinity in the supergravity calculation. However, it
would be nice to see this explicitly, for example by doing the calculation on
the Taub-NUT space instead of the ALE space.

The second open problem stems from the fact that the classical and quantum parts
of \ref{eq:fullresult} satisfy conjugate ``holomorphic anomaly equations''
\cite{bcov93}, i.\ e.\ they are not truly holomorphic quantities in $z$, but 
they are holomorphic in a background independent sense as explained by Witten 
in \cite{witten93}. Integrating over the C-field fluxes $x$ would give a truly
holomorphic result. It would be nice to understand these facts and the relation
to topological theories better.

\begin{acknowledgments}
The work reported here was done in collaboration with R. Dijkgraaf and E. 
Verlinde \cite{dvv02}. I would like to thank the organizers of the Carg\`ese 
summer school for the opportunity to present this work. 
\end{acknowledgments}


\begin{chapthebibliography}{99}

\bibitem{dvv02}
R.\ Dijkgraaf, E.\ Verlinde and M.\ Vonk, {\em On the partition sum of the NS
five-brane,} hep-th/0205281.

\bibitem{gk99}
A.\ Giveon and D.\ Kutasov, {\em Little string theory in a double scaling
limit,} JHEP 9910 (1999), hep-th/9909110.

\bibitem{witten96}
E.\ Witten, {\em Five-brane effective action in M-theory}, J. Geom. Phys. {\bf 
22} (1997) 103, hep-th/9610234.

\bibitem{ov96}
H.\ Ooguri and C.\ Vafa, {\em Two-dimensional black hole and singularities of CY
manifolds}, \np{463}{1996}{55}, hep-th/9511164.

\bibitem{agnt94}
A.\ Antoniadis, E.\ Gava, K.\ S.\ Narain and T.\ R.\ Taylor, {\em Topological
amplitudes in string theory}, \np{413}{1994}{162}, hep-th/9307158.

\bibitem{bcov93}
M.\ Bershadsky, S.\ Cecotti, H.\ Ooguri and C.\ Vafa, {\em Holomorphic anomalies
in topological field theories,} \np{405}{1993}{279}, hep-th/9302103.

\bibitem{witten93}
E.\ Witten, {\em Quantum background independence in string theory}, in Salamfest
1993, 257, hep-th/9306122.

\end{chapthebibliography}
\end{document}